\documentclass[prb,twocolumn,showpacs]{revtex4}
\usepackage{graphicx}

\begin{document}

\title{ Variable-range-hopping conductivity of half-doped bilayer manganite
LaSr$_{2}$Mn$_{2}$O$_{7}$ }

\author{X. J. Chen$^{1}$, C. L. Zhang$^{1}$, J. S. Gardner$^{2}$, J. L. Sarrao$^{3}$,
and C. C. Almasan$^{1}$ }
\affiliation{
$^{1}$Department of Physics, Kent State University, Kent, OH 44242\\
$^{2}$NRC-NPMR, Chalk River Laboratory, Chalk River ON KOJ 1PO, Canada\\
$^{3}$MST-10, Los Alamos National Laboratory, Los Alamos, NM 87545 }
\date{\today}

\begin{abstract}
We report measurements of in-plane $\rho_{ab}$ and out-of-plane $\rho_{c}$ resistivities 
on a single crystal of the half-doped bilayer manganite LaSr$_{2}$Mn$_{2}$O$_{7}$. In the
temperature $T$ range 220 to 300 K, the resistive anisotropy $\rho _{c}/\rho _{ab}=A+B/T$ 
($A$ and $B$ constants), which provides evidence for the variable-range-hopping
conduction in the presence of a Coulomb gap. This hopping mechanism also accounts for the
quadratic magnetic field $H$ and $\sin^{2}\varphi$ dependences of the negative 
magnetoresistivity $\ln \left[\rho_{i}(T,H,\varphi)/\rho_{i}(T,H=0)\right]$ ($i=ab,c$), where
$\varphi$ is the in-plane angle between the magnetic field and the current.
\end{abstract}

\maketitle

Since the discovery of colossal magnetoresistance in manganese oxides, much effort has been
devoted to understand their magnetic and electrical transport properties. It has been shown
\cite{coey,font,viret} that the temperature $T$ dependence of the electrical conductivity in
the paramagnetic phase is well described by Mott variable-range-hopping (VRH),
\cite{mott69,ambeg} that is,
\begin{equation}
\label{mot}
\sigma = \sigma _{0}\exp\left[-\left(\frac{4\nu _{c}\alpha
^{d}}{k_{B}N(E)T}\right)^{p}\right]~~,
\end{equation}
Here $p=1/(d+1)$, with $d$ the dimensionality, $\sigma_{0}$ is a constant which depends on
the assumptions made about electron-phonon interaction, $\nu_{c}$ is a dimensionless constant,
$\alpha $ is the reciprocal of the localization length $\xi$, and $N(E)$ is the density of
states at the Fermi level. However, Eq. (\ref{mot}) usually yields a small value for
the localization length ($\xi<0.2$ nm) of the cubic manganites when $N(E)$ is deduced from
the electronic heat capacity coefficient $\gamma$.\cite{coey} Since $\xi$ is expected to be
of the order of the Mn-Mn distance, such a small $\xi$ is incompatible with conventional VRH
and has an unphysical meaning.

In order to address this inconsistency, Viret, Ranno, and Coey \cite{viret} developed a VRH 
model based on the idea of magnetic localization. Although the estimated value of $\xi$ is
physically plausible in this case, it strongly depends on the splitting energy $U_{m}$
between spin-up and spin-down $e_{g}$ bands. At present, it is highly desirable to perform
accurate measurements of $U_{m}$ for manganese oxides.

The derivation of $dc$ conductivity as given by Eq. (\ref{mot}) is based upon the assumption 
that the density of states near the Fermi level is constant. Efros and Shklovskii developed a
VRH theory which takes into account the electron-electron Coulomb interaction, which reduces 
the density of states near the Fermi level.\cite{efros,shkl} It was suggested that the
Coulomb interaction may have an important effect on the hopping conduction of electrons in 
manganese oxides.\cite{varma,shen} Hence, the theory of weak localization and VRH in the
presence of a Coulomb gap, as developed by Shklovskii and Efros (SE), could account for the 
temperature dependence of conductivity in manganites. Specifically, for half-doped
manganites, the Coulomb interaction is believed to be not only the source of charge 
ordering,\cite{shen2} but also the convincing candidate for the anisotropy in the
orbital-ordered states.\cite{ishi} Therefore, half-doped manganites can be model systems for 
clarifying whether the SE-VRH conduction mechanism dominates the electrical transport in
their paramagnetic state.

In this paper, we present in-plane $\rho_{ab}$ and out-of-plane $\rho_{c}$ resistivity
measurements of a half-doped LaSr$_{2}$Mn$_{2}$O$_{7}$ single crystal as a function of
temperature, magnetic field $H$, and the in-plane angle $\varphi$ between the magnetic field 
and the electrical current. Both resistivities follow well a VRH behavior for $220 \leq T
\leq 300$ K. However, as shown before for the cuprates, the temperature dependence of the
resistive anisotropy $\rho_{c}/\rho_{ab}$ in the VRH regime is a much more effective
indicator of the type of hopping than the traditional method based on Eq. (1).\cite{levi}  
Here, we show that $\rho _{c}/\rho _{ab}=A+B/T$ for $220 \leq T \leq 300$ K, which
unambiguously indicates VRH in the presence of a Coulomb gap. This hopping mechanism also 
accounts for the $H$ and $\varphi$ dependences of the  magnetoresistivities $\ln
\left[\rho_{i}(T,H,\varphi)/\rho_{i}(T,H=0)\right]$ ($i=ab,c$). We also demonstrate that the 
negative magnetoresistivity in the VRH regime is a result of the increase of the
localization length, hence, the decrease of resistivity, when a magnetic field is applied.

Measurements of $\rho_{ab,c}(T,H,\varphi)$ of a single crystal of LaSr$_{2}$Mn$_{2}$O$_{7}$ 
were performed using a multiterminal lead configuration,\cite{jiang} over a temperature range
from 2 to 300 K and in magnetic fields up to 14 T. The crystal was cleaved from a boule 
prepared by the optical floating-zone method, as reported elsewhere.\cite{argy} A total of
eight low-resistance electrodes were applied on the top and bottom faces of the crystal 
using thermally treated silver paint. The electrical current was always applied along one
of the crystal faces, while the top and bottom face voltages were measured simultaneously. 
The rotation of the sample was performed along the $c$ direction, keeping the applied
magnetic field within the MnO$_{2}$ planes. The angle $\varphi$ is defined to be $0^{0}$ 
($90^{0}$) when the magnetic field is parallel (perpendicular) to the current. The $dc$
magnetization measurements were carried out using a superconducting quantum interference 
device (SQUID) magnetometer.

\begin{figure}[t]
\begin{center}
\includegraphics[width=\columnwidth]{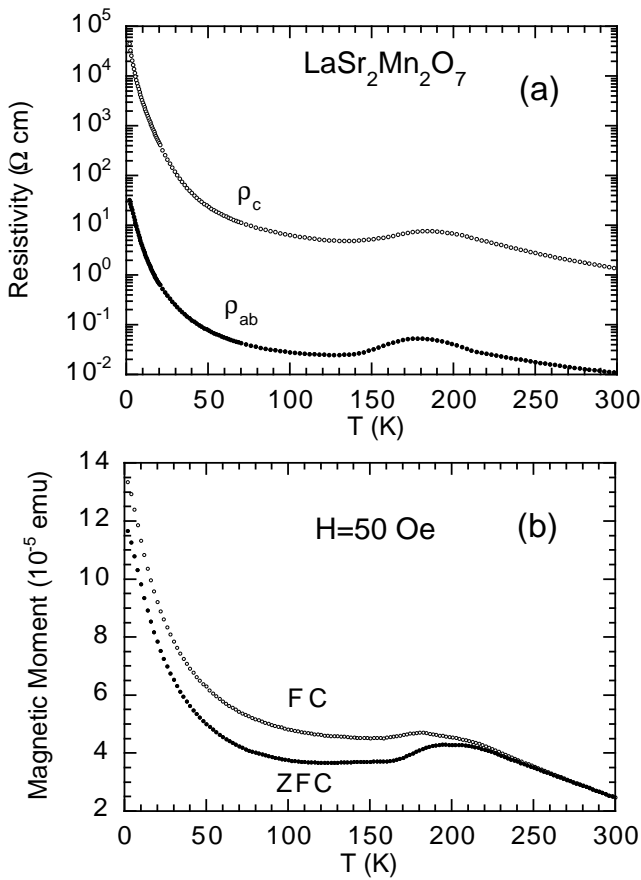}
\end{center}
\caption{Temperature dependence of (a) in-plane $\rho_{ab}$ and out-of-plane $\rho_{c}$
resistivities measured in zero field and (b) magnetization $M$ measured both in increasing 
the temperature after cooling the sample in zero field to 2 K (ZFC measurement), and in
decreasing the temperature in the presence of an applied magnetic field (FC measurement) 
of LaSr$_{2}$Mn$_{2}$O$_{7}$. }
\label{fig1}
\end{figure}

The temperature dependences of the zero-field resistivities $\rho_{ab}$ and $\rho_{c}$, and
of the zero-field-cooled (ZFC) and field-cooled (FC) magnetization $M$ measured in an applied 
magnetic field of 50 Oe with $H||c$ are shown in Fig. 1(a) and 1(b), respectively. These
plots display several features which correlate with the charge-ordering and antiferromagnetic 
transitions in this half-doped compound. A steep increase of resistivity as well as a
hysteresis in magnetization is observed just below 220 K, signaling the presence of a charge 
and orbital ordered phase. Ordering of the $d_{3x^{2}-r^{2}}/d_{3y^{2}-r^{2}}$ orbitals of
the Mn$^{3+}$ ions, resulting from a cooperative Jahn-Teller distortion, accompanied by a 
real space ordering in the Mn$^{3+}$/Mn$^{4+}$ distribution for $T<$ 220 K has been confirmed
by electron, neutron, and x-ray diffractions.\cite{argy,jqli,tkim,mkub,tcha,ywak} A maximum 
near 180 K visible in all resistivity and magnetization curves coincides with the onset of
antiferromagnetism, while a broad minimum around 100 K corresponds to the transition to
a canted spin state. These temperature values are consistent with neutron diffraction data.
\cite{argy,mkub}

In half-doped manganites, the Coulomb interaction modifies the density of states at the Fermi
level\cite{Medve} and would affect the charge transport. According to the SE-VRH theory,
\cite{shkl} the temperature dependence of the resistivity in the VRH regime is given by:
\begin{equation}
\label{vrh}
\rho =\rho _{0}\exp \left[\left(\frac{T_{0}}{T}\right)^{1/2}\right]~~,
\end{equation}
where $\rho_0$ is a constant and $T_{0}=2.8e^{2}/(4\pi k_{B} \epsilon_{0} \xi)$. Here,
considering the high-density of electrons in manganites, we take the background dielectric 
constant $\kappa =1$ like in the jullium model for simple metals.\cite{shen}

Figure 2(a) shows semilog plots of zero-field $\rho_{ab,c}$ vs $T^{-1/2}$. Clearly, both
resistivities exhibit VRH in the paramagnetic state (temperature range 220 to 300 K) above 
the charge-ordering transition temperature. However, over such a narrow $T$ range, one cannot
reliably distinguish between a two-dimensional (2D) Mott-VRH ($p=1/3$ in Eq. (1)) and a 
SE-VRH ($p=1/2$). Moreover, the parameter $T_{0}$ determined by fitting the data in Fig. 2(a)
with Eq. (2) is larger for $\rho_{c}$ than for $\rho_{ab}$ by approximately a factor of 2. As 
shown below, this is the result of ignoring the temperature dependence of the
pre-exponential factor. In fact, $T_{0}$ turns out to be the same for $\rho_{ab}$ and
$\rho_{c}$, in agreement with the theory of anisotropic hopping.\cite{shkl}

\begin{figure}[t]
\begin{center}
\includegraphics[width=\columnwidth]{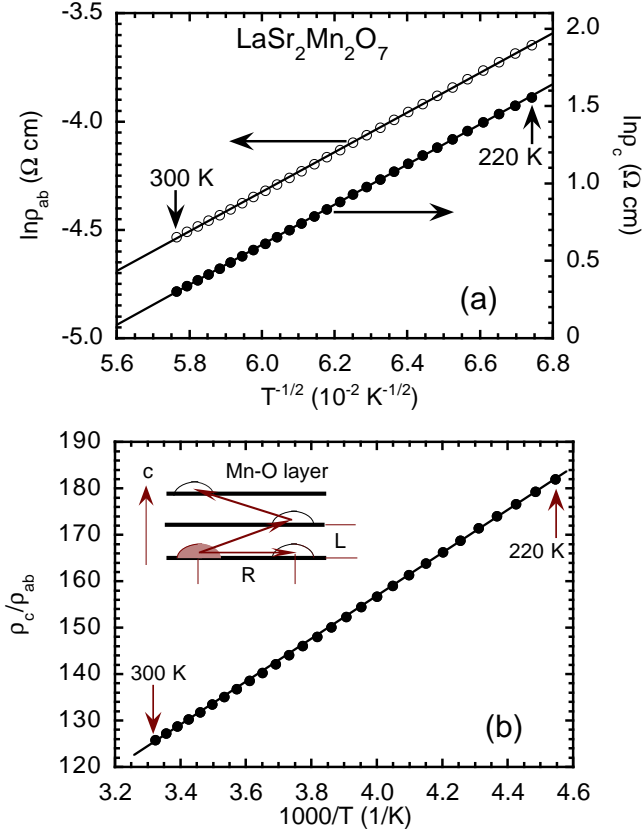}
\end{center}
\caption{ (a) Semilog plot of the zero-field resistivities $\rho_{i}$ ($i=ab,c$) vs $T^{-1/2}$. 
The straight lines are guides to the eye. (b) Resistive anisotropy $\rho_{c}/\rho_{ab}$
vs 1000/T. The straight line is a linear fit of the data with Eq. (3). Inset: Hopping
processes in the Mn-O bilayer and the zigzag path in the $c$ direction. }
\label{fig2}
\end{figure}

A plot of the resistive anisotropy $\rho_{c}/\rho_{ab}$ vs 1000/T for LaSr$_{2}$Mn$_{2}$O$_{7}$,
displayed in Fig. 2(b), clearly shows that, in the temperature regime where both 
resistivities follow the VRH model, there is the following relationship between resistivities:
\begin{equation}
\rho_{c}=(A+\frac{B}{T})\rho_{ab}~~,
\end{equation}
with $A=-28.78$ and $B=4.64\times 10^{4}$ K. It has been shown that the resistive anisotropy
of an anisotropic material is given by:\cite{levi}
\begin{equation}
\frac{\rho_{c}}{\rho_{ab}}=\frac{1}{2}\frac{<R^{2}P_{ab}(R)>}{L^{2}<P_{c}(R)>}
\approx \frac{1}{2}\frac{<R^{2}><P_{ab}(R)>}{L^{2}<P_{c}(R)>}~~,
\end{equation}
where $R$ is the in-plane hopping distance, $L$ is the distance between adjacent bilayers,
$P_{ab}$ is the hopping probability between two states on the same bilayer separated by a 
distance $R$, and $P_{c}$ is the hopping probability between two states located on adjacent
bilayers and separated by a {\it variable lateral} distance $R$ and {\it fixed} transverse
distance $L$ (see inset to Fig. 2(b)). Equation (4) reflects the experimental relationship
given by Eq. (3) if $P_{ab}(R) \propto P_c(R)$. Then, $\rho_{c}/\rho_{ab} \propto 
<R^{2}>/L^{2}$. This implies that the experimentally observed $T^{-1}$ dependence of the
anisotropy,  given by Eq. (3), is a result of increasing mean square in-plane hopping 
distance with decreasing temperature as $<R^2> \propto A+BT^{-1}$ and $T$ independent
out-of-plane step $L$. In the SE-VRH model, the Coulomb interaction leads to an increase of 
the average in-plane hopping distance with decreasing temperature as
$<R>=(\xi/4)(T_0/T)^{1/2}$.\cite{cast} Thus, the $T$ dependence of the anisotropy given by  
Eq. (3) unambiguously points toward SE-VRH as the hopping conduction mechanism for
$220 \leq T \leq 300$ K. 

On the other hand, in the case of the 2D Mott-VRH conduction, the average in-plane hopping 
length $<R> \propto \xi(T_{0}/T)^{1/3}$,\cite{shkl} which gives $\rho_{c}/\rho_{ab}\propto
<R^{2}>/L^{2}\propto T^{-2/3}$. Such a resistive anisotropy has been found in the insulating
PrBa$_{2}$Cu$_{3}$O$_{7-\delta}$,\cite{levi} but not in the present bilayer manganite. Therefore,
although the resistivity data of the present bilayer manganite can be fitted with Eq. (1)
with $p=1/3$ almost as well as with $p=1/2$ (SE-VRH), the $T$ dependence of the resistive
anisotropy excludes the Mott-VRH conduction and conclusively points toward SE-VRH conduction.

\begin{figure}[t]
\begin{center}
\includegraphics[width=\columnwidth]{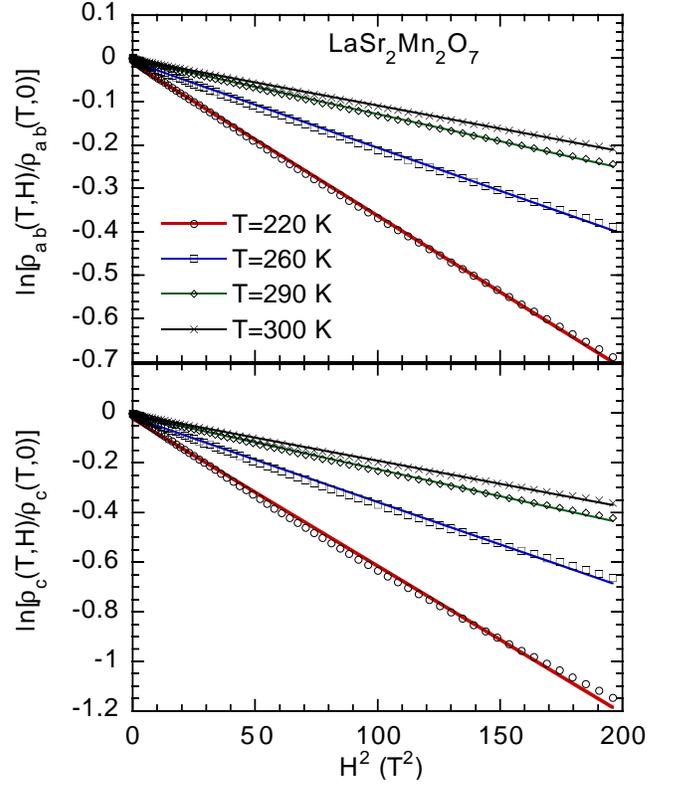}
\end{center}
\caption{ Magnetoresistivities $\ln \rho_{i}(T,H)/\rho_{i}(T,H=0)$ ($i=ab,c$) vs $H^{2}$ of
LaSr$_{2}$Mn$_{2}$O$_{7}$ for various temperatures in the variable-range-hopping regime. The
straight lines are guides to the eye. }
\label{fig3}
\end{figure} 

Equation (3) also indicates that, when one takes into account the pre-exponential factor  
$A+BT^{-1}$ in $\rho_c(T)$, both resistivities have the {\it same} exponential factor
$exp[(T_0/T)^{1/2}]$. Using this experimentally determined $T_0$ of 0.71 eV, we get a
localization length $\xi_{ab}$ = 56.8 $\AA$. This value is about 15 times larger than the Mn-Mn
separation of 3.87 $\AA$  in this half-doped bilayer manganite.\cite{argy,mkub} Thus, the
localization length obtained based on SE-VRH conduction is physically reasonable. Also, as
discussed above, $\xi_{c} \equiv L \approx$ 4 $\AA$. This indicates that the charge transport in this 
manganite is 2D in nature, with the in-plane localization length one order of magnitude larger than
the out-of-plane one. As a note, VRH theory in the presence of a Coulomb gap leads to the same 
temperature dependence of the resistivity (Eq. (2)) for both 2D and 3D cases.

For the SE-VRH model to be valid, the average hopping energy $\Delta$ should be equal to the
energy $U$ of the Coulomb interaction between the sites. $\Delta$ and $U$ can be estimated
from the experimental data as follows. The transition from nearest-neighbor-hopping NNH to
SE-VRH takes place at a critical temperature $T_{V}$ at which the NNH energy $E_{A}$
[$\rho(T)=\rho_0exp(E_A/k_BT$)] becomes equal to the average SE-VRH energy $\Delta$; i.e.,
$\Delta \equiv E_{A}=k_{B}T(T_{0}/T)^{1/2}|_{T=T_{V}}$. The high-temperature electrical 
transport of bilayer manganites is usually described by nearest-neighbor thermally-activated
hopping.\cite{chen,mori} Also, since the resistivity data follow well the SE-VRH mechanism 
up to the highest measured temperature of 300 K, we take $T_{V}=300$ K as the crossover
temperature from NNH for $T>300$ K to SE-VRH for $T<300$ K. With the experimentally determined
$T_{0}=0.71$ eV, we obtain $\Delta = k_{B}\sqrt{T_{0}T_{V}}=0.136$ eV. We next determine $U$
from $U\approx e^{2} /(4 \pi \epsilon_{0} \bar{R_{0}})$, where the background dielectric 
constant is again taken to be $\kappa =1$. The $T$-independent average distance $\bar{R_{0}}$
between hopping sites can be determined from the simple formula 
$n^{-1}=(4\pi/3)(\bar{R_{0}}/2)^{3}$, where $n$ is the carrier density given by $n=1/eR_{H}$,
with $R_{H}$ the Hall coefficient. Taking $R_{H}=4.0\times 10^{-4}$ cm$^{3}$/C for half-doped 
manganites,\cite{asam} we obtain $\bar{R_{0}}=1.07\times 10^{-8}$ m. The substitution of this
estimated $\bar{R_{0}}$ into the equation for $U$ yields $U=0.134$ eV, in excellent agreement 
with $\Delta = 0.136$ eV, determined above. This result further confirms that SE-VRH model
provides a consistent description of the charge transport for $220\leq T\leq 300$ K. 

\begin{figure}[t]
\begin{center}
\includegraphics[width=\columnwidth]{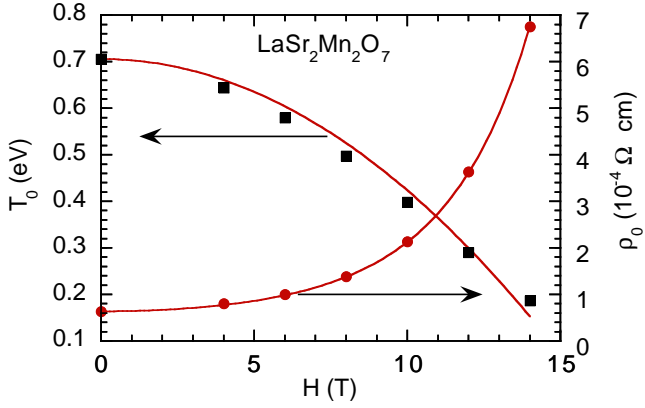}
\end{center}
\caption{ Magnetic field $H$ dependence of the parameters $T_{0}$ and $\rho_{0}$ in Eq.
(\ref{vrh}) obtained by fitting in-plane resistivity data measured for $220 \leq T \leq 300$ 
K and at various magnetic fields of LaSr$_{2}$Mn$_{2}$O$_{7}$. }
\label{fig4}
\end{figure}

Next, we show that the negative magnetoresistivity $\ln [\rho_{i}(T,H)/\rho_{i}(T,0)]$ 
($i=ab,c$) and its magnetic field dependence observed in LaSr$_{2}$Mn$_{2}$O$_{7}$ for $220
\leq T \leq 300$ K can also be understood based on the SE-VRH model. In Altshuler, Aronov, 
and Khmelnitskii (AAK) localization theory,\cite{blal} the effect of an applied magnetic
field is to increase the localization length $\xi$, which decreases the resistivity and gives 
rise to negative magnetoresistivity.  AAK obtained the following expression for the negative
magnetoresistivity in the SE-VRH regime:\cite{blal}
\begin{equation}
\label{orb}
\ln \frac{\rho(T,H)}{\rho(T,0)}=-C\left(\frac{ea^{2}H}{\hbar c}\right)^{1/2\nu}\ln
\frac{\rho(T)}{\rho_{0}}~~,
\end{equation}
where $C$ is a positive constant and $\nu$ is the critical index for the localization radius
and conductivity in the scaling theory of the metal-insulator transition. They predicted
$\nu=1/4$ and $1$ for weak and strong magnetic fields, respectively.

The magnetic field dependence of $\ln [\rho_{i}(T,H)/\rho_{i}(T,0)]$ ($i=ab,c$) for $T\geq 
220$ K is shown in Fig. 3. The magnetoresistivities are, indeed, negative and follow
a $H^{2}$ dependence in magnetic fields up to 14 T. This implies that $\nu=1/4$ in Eq. (5). 
Therefore, the magnetic-field dependence of the magnetoresistivity data of
LaSr$_{2}$Mn$_{2}$O$_{7}$ is consistent with the AAK theoretical prediction of the SE-VRH 
for weak magnetic fields.

\begin{figure}[t]
\begin{center}
\includegraphics[width=\columnwidth]{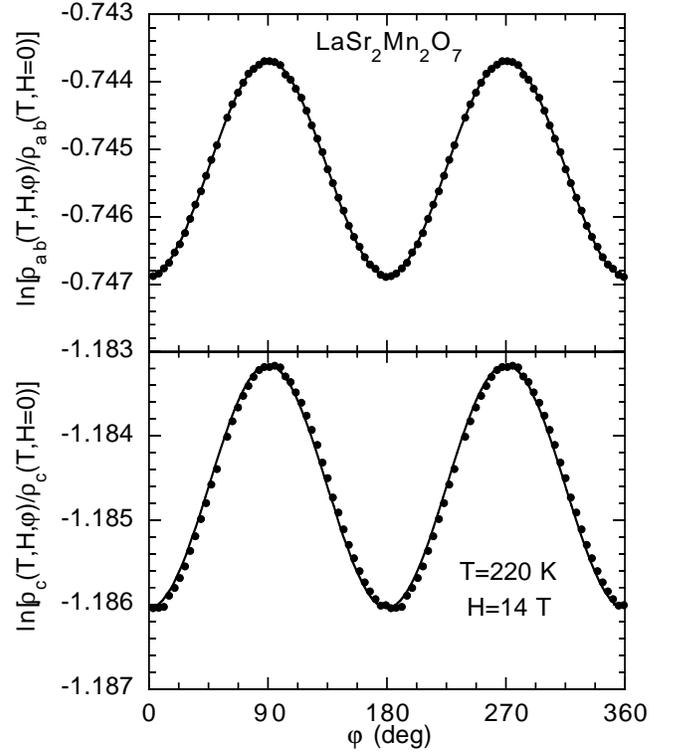}
\end{center}
\caption{Magnetoresistivities $\ln \rho_{i}(T,H,\varphi)/\rho_{i}(T,H=0)$ ($i=ab,c$) vs the 
angle $\varphi$  between the applied magnetic field and the current of
LaSr$_{2}$Mn$_{2}$O$_{7}$ measured at $T=220$ K and in $H=14$ T. The lines are fits of the 
data with Eq. (6). }
\label{fig5}
\end{figure}

We also measured the temperature dependence of the resistivities at various magnetic fields.
$T_{0}$ and $\rho _{0}$ are then determined by fitting the $\rho_{ab}$ vs $T$ curves in the 
temperature range 220 to 300 K with Eq. (2). Figure 4 shows the magnetic field dependence of
$T_{0}$ and $\rho _{0}$. Notice that the effect of an applied magnetic field is to decrease 
$T_{0}$ and to increase $\rho _{0}$. The former is quadratic in field; $i.e.$,
$T_{0}(H)=T_{0}[1-\beta H^{2}]$, with $T_{0}=$ 0.71 eV and the fitting coefficient $\beta=$ 
4.00$\times 10^{-3}$ T$^{-2}$. The prefactor $\rho_{0}$ is well described by
$\rho_{0}(H)=\rho_{0}\exp(\eta H^{2})$, with $\rho_0 = 6.36\times 10^{-5}$ $\Omega$ cm and 
$\eta=$ 1.21$\times 10^{-2}$ T$^{-2}$. Therefore, the negative magnetoresistivity of this
bilayer manganite is a result of the decrease of $T_{0}$ and the increase of the prefactor 
$\rho_{0}$ of the SE-VRH resistivity when a magnetic field is applied. Noting that
$T_{0}\propto 1/\xi$, the application of a magnetic field, indeed, gives rise to an increase 
in $\xi$. Since $\rho_{0}$ is an increasing function of magnetic field, the AAK localization
theory\cite{blal} implies $\rho_{0}\propto \xi$. This behavior can be understood within the 
Landauer expression\cite{land} $\rho=(2h/e^{2})\xi$ by neglecting the inelastic scattering in
the zero-temperature limit.  

The anisotropy of magnetoresistivity in heavily doped semiconductors has provided compelling 
evidence for the localization theory responsible for the negative magnetoresistivity.
\cite{blal,kawa} The anisotropy of both magnetoresistivities of 
LaSr$_{2}$Mn$_{2}$O$_{7}$ is shown in Fig. 5, which is a plot of in-plane and out-of-plane
magnetoresistivity, measured at 220 K in a magnetic field  of 14 T, vs the angle between the 
magnetic field and the current. Both magnetoresistivities vary as $\sin^2 \varphi$ when the
magnetic field is rotated in the MnO$_{2}$ plane, with a maximum value when $H\perp I$ and a 
minimum one when $H\parallel I$. These data are well fitted with
\begin{equation}
\label{ang}
\ln \frac{\rho (T,H,\varphi)}{\rho (T,0)}=\ln \frac{\rho (T,H,0)}{\rho (T,0)}
\left(1+P\sin^{2}\varphi \right)^{1/4\nu}~~,
\end{equation}
with $\nu=1/4$ and the only fitting parameter $P=4.27\times 10^{-3}$ and $2.43\times 10^{-3}$
for the in-plane and out-of-plane resistivity, respectively. Equation (6) is consistent with 
the prediction of the localization theory in the SE-VRH regime for weak magnetic fields
($\nu=1/4$), in which $P=(D_{\parallel}-D_{\perp})/D_{\perp}$, with $D_{\parallel}$ and 
$D_{\perp}$ the diffusion coefficient parallel and perpendicular to the current, respectively.
\cite{blal} Assuming that the current is applied along the crystallographic direction $a$,
$D_{\parallel}=2E_{F}\tau/3(m_{a}^{3}/m_{b}^{2})$ and $D_{\perp}=2E_{F}\tau/3m_{b}$.\cite{kawa}
Since the relaxation time $\tau$ of an electron is isotropic and the components of the 
effective mass tensor are almost equal along the $a$ and $b$ directions, one would expect a
small difference between $D_{\parallel}$ and $D_{\perp}$. Thus, the small anisotropic 
magnetoresistivity, i.e., small $P$ value, is the result of a small difference between the
in-plane diffusion coefficients.   

In conclusion, we report in-plane and out-of-plane magnetoresistivity measurements performed 
on LaSr$_{2}$Mn$_{2}$O$_{7}$, a half-doped bilayer manganite. The resistivity clearly follows
a variable-range-hopping behavior for $220 \leq T \leq 300$ K. However, due to this narrow 
$T$ region, one cannot conclusively determine the type of hopping conduction from the
resistivity data. Nevertheless, the $T$ dependence of the resistive anisotropy 
($\rho _{c}/\rho _{ab}=A+B/T$) indicates that the hopping conduction in this $T$ range is of
SE-type, i.e., takes place in the presence of a Coulomb gap. The determined localization 
length $\xi = 56.8 \AA$, average hopping energy $\Delta =0.136$ eV, and energy of the Coulomb
interaction $U=0.134$ eV have physically reasonable values. In magnetic fields up to 14 T, 
the magnetoresistivity $\ln [\rho_i(T,H,\varphi)/\rho_i(T,0)]$ ($i=ab,c$) is negative and its
magnitude increases proportional to $H^{2}$ and $\sin^{2}\varphi $. These results provide 
convincing evidence of the SE-type variable-range-hopping conductivity in half-doped manganites.

This research was supported at KSU by the National Science Foundation under Grant No. DMR-0102415.
The work at LANL was performed under the auspices of the U.S. Department of Energy.

\clearpage

\end{document}